\documentclass[floatfix,aps,prb,twocolumn,superscriptaddress,showpacs]{revtex4}
\usepackage{graphicx}
\usepackage{epsfig}
\usepackage{dcolumn}
\usepackage{bm}

\begin{document}

\title{Edge States in Doped Antiferromagnetic Nano-Structures}

\author{A. L. Chernyshev}
\affiliation{Department of Physics, University of California, Irvine,
  California 92697} 
\author{A. H. Castro Neto}
\affiliation{Department of Physics, Boston University, Boston,
  Massachusetts 02215}
\author{S. R. White}
\affiliation{Department of Physics, University of California, Irvine,
  California 92697} 

\date{\today}

\begin{abstract}
We study competition between different phases in
%discuss a phase diagram of 
a strongly correlated nano-structure with an edge.
Making use of the self-consistent Green's function and
density matrix renormalization group methods, we study
a system 
%nano-structure 
described by the $t$-$J_z$ and $t$-$J$ models 
on a strip of a square lattice with a 
%controllable 
linear hole density $n_{||}$.   
At %a lower hole density and at 
intermediate interaction strength $J/t$  we find edge stripe-like states, 
reminiscent of the bulk stripes that occur at 
%higher densities and 
smaller $J/t$. We find that %due to collective effects
stripes attach to edges more readily than 
%single holes or 
hole
pairs, and
that the edge stripes 
can exhibit a peculiar phase separation.
%are stable at lower linear hole density than bulk stripes.
\end{abstract}

\pacs{71.10.Fd, 74.72.-h, 71.10.Pm}

\maketitle

It is well established by now that in the ground state of the
two-dimensional (2D)
strongly correlated cuprates, much studied in the context of the
high-T$_c$ problem,\cite{tranquada,dmrg} charge carriers tend to
be distributed inhomogeneously. Such inhomogeneities occur as an
attempt to reduce the frustration between the exchange and kinetic
energies arising from the antiferromagnetic (AF) 
background.\cite{tranquada}
It is conceivable that such frustration can be further reduced at 
a boundary. This would induce an effective 
attraction of the bulk inhomogeneity to the edge, thus 
leading to a novel, many-body type edge state. 
A bulk doped Mott insulator might have various types of striped,
paired, or unpaired ground states. Ordinarily one is not concerned
with surface states. 
However, for cuprate nanoscale structures, 
which are of interest due to recent progress in 
nano-synthesis,\cite{dale} one may potentially 
have a system with a finite doping
in which the holes are bound to the edges.
In this case, understanding the possible edge states is crucial. 

Here, we consider
the simplified case of a half-plane, or similarly, one end of a long open
cylinder, with a vanishingly small doping sufficient to form a finite
{\it linear} concentration $n_\parallel=N_h/L$ of holes near the edge.
$N_h$ is the number of holes and $L$ is the linear size of the
system. 
Given this configuration, 
several questions arise. For example, if the holes form a single 
stripe in the bulk, does this stripe attach itself to the edge? Do single
pairs attach to the edge? Can one have a one-dimensional (1D) 
``edge phase separation'' with the holes
clustered together in edge droplets? 
%What is the nature of the excitations of
%the various possible edge phases?

In this paper we address some of 
these questions in the context of the $t$-$J_z$ and $t$-$J$ models,  
using the self-consistent Green's function and the density
matrix renormalization group (DMRG) methods.
We obtain an approximate edge phase diagram in which the linear hole
density $n_\parallel$ is a control parameter. 
We find analytically, for the 
$t$-$J_z$ model, that the stripe 
edge states do exist in a substantial region of
the phase diagram.
%----------------------------------------------------------------
\begin{figure}[t] 
\includegraphics[angle=0,width=8cm]{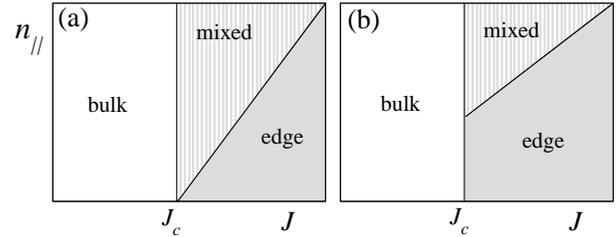}
\caption{Phase diagram of a system with an edge, in
the linear hole concentration ($n_\| $) v.s.
attraction to the edge ($J$) plane. 
(a) band-like case, (b) case of complex bulk/edge states. 
The labels of the phases indicate where the holes reside.}
\label{phd}
\end{figure} 
%----------------------------------------------------------------
This is verified for both the $t$-$J_z$ and $t$-$J$ models using 
DMRG for systems up to
$11\times 8$ sites with periodic/open and $20\times 5$ with open
boundary conditions. We also find that the Ising anisotropy enhances
tendency to the edge states formation.
A remarkable feature of these states is that they are
reminiscent of and, as we show in some detail, are
directly related to the bulk stripe states observed in the
studies of strongly correlated models as well as 
in real high-T$_c$ materials.\cite{tranquada,dmrg,sasha}
We argue that such edge states can be a common feature of
strongly correlated nano-systems. 
 
With our  Fig. \ref{phd} we outline two generic possibilities 
for the formation of the edge state as a function of attraction to the
boundary (controlled by $J$) and linear hole concentration $n_\|$. 
Fig. \ref{phd}(a) corresponds to the case when both the 2D bulk and the
1D edge state are band-like. At small $J$ the kinetic
energy in the bulk is lower and the holes occupy the bulk states only
(bulk). At $J=J_c$ the bottom of the edge-band becomes 
degenerate with the bulk band. As the attraction to the
edge is increased further, a fraction of holes from the bulk 
populates the edge band (mixed). Upon the further increase of $J$ all the
holes are in the edge band (edge). Note that in the bulk and mixed states 
the chemical potential is pinned at the bottom of the bulk band as the
finite linear concentration $n_\|=N_h/L$ corresponds to zero bulk
concentration $x=N_h/L^2$, as $L\rightarrow\infty$. 
Fig. \ref{phd}(b) outlines a different generic 
scenario in which the edge state formed at $J=J_c$ accumulates a
finite linear concentration of holes. That is,
at $J_c$ a finite amount of holes $n_\|^c$ becomes attached to the 
boundary at once. This could occur, for example, if the lowest energy
state is a stripe attached to the edge.

In this work we focus on a particular case of the strongly correlated
models, the $t$-$J_z$ and the $t$-$J$ models, and,
therefore, on a particular case of such a frustration characteristic
to the doped antiferromagnets. 
We find, by means of analytical and numerical approaches, that the edge 
states do form in these systems and that their phase diagram belongs 
to the class described in Fig. \ref{phd}(b). 

First, we provide a general reasoning for the edge states to occur in the
$t$-$J$-like models.  Excitations in such models doped with 
few holes are described as spin-polarons, which either form moderately
bound pairs or move independently.\cite{sasha} 
It is  clear that at some $J/t$ such excitations would 
reside close to an edge where they frustrate the AF background
less. Thus, edge holes or edge pairs can be formed. 
At larger densities the holes are known to
be in the ``bulk'' quasi-1D stripes 
%in a wide range of $J/t$  
with internal hole concentration close to $n_\|=1/2$.\cite{sasha} 
Although the kinetic energy advantage for holes in the stripe is
substantial, the magnetic energy cost of the antiphase-domain wall (ADW)
associated with the stripe is considerable too, $E_{mag}\sim J$ per
hole. However, this magnetic energy of the domain wall can be
significantly reduced at the edge. Therefore, 
an edge stripe can be formed.
One may also expect that the low-density stripes will be 
attracted to the edge more readily because their magnetic energy
is higher ($E_{mag}\sim J/n_\|$)
and its reduction should be more effective for them.
Thus,  one expects that in some range of $n_\parallel$ and $J$
an edge stripe can be formed. 

Note that within this scenario edge stripes constitute
many-body states formed due to a non-trivial collective effect.
That is, the energy is reduced  
due to attraction of the stripe as an entity.
This means that if
the ground state of the $t$-$J$ model would be given by a dilute gas of 
spin polarons then the physics outlined in Fig. \ref{phd}(a) should be 
applicable to the bulk and edge bands of these excitations. However, 
if the ground state is a stripe, then we expect the edge phase diagram 
of Fig. \ref{phd}(b).

Another important question here is: what is the ground state of the 
$t$-$J$ model in the limit of small $n_\|\ll 1/2$? 
The most likely answer 
 is that the ground state is given by a closed 
``stripe loop'' with an AF antiphase shift across the loop and internal hole 
concentration close to the ``optimal'' $n_\parallel = 1/2$, as supported by
the observation in our previous work, Ref. \onlinecite{sasha}.
Such loops can be seen as nucleations of the ``straight'' stripes. 
In this regime we expect the edge states to consist of short loop-like
segments.
%It is clear that the edge states in this range of doping should represent 
%counterparts to such stripe loops.

To demonstrate the described tendencies explicitly and quantitatively
we consider the $t$-$J$ model:
\begin{eqnarray}
\label{tj}
{\cal H}= -t\sum_{
%\langle 
ij, 
%\rangle
\sigma} 
%\left( 
\tilde{c}^{\dag}_{i\sigma}
\tilde{c}^{\phantom{\dag}}_{j\sigma}
%+\mbox{h.c.}
%\right)
%\\
%\nonumber
%&&
+ \frac{J}{2} \sum_{
%\langle 
ij 
%\rangle
}
\left(
{\bf S}_i \cdot {\bf S}_j -
  \frac{n_i n_j}{4} 
\right)
,
\end{eqnarray}
 on a strip of a square lattice and 
in the standard notations of constrained fermion and spin operators,
$J$ is the superexchange and $t$ is the kinetic energy, and summation is
over nearest-neighbor sites.

First, we consider a simpler problem where we impose a local constraint
on the linear hole density. Namely, a given $n_\parallel$ will be assumed to be
distributed homogeneously along the boundary. 
This consideration will help to clarify the details of the competition 
between different bulk and edge states such as a dilute gas of holes/pairs and 
``straight'' stripes, but will exclude the 1D phase separated states such as 
loops.  In reality,
such a constraint may be imposed by the confinement effect of
finite sizes of a structure or by extra interactions beyond the
considered models. In a nano-size system single-hole and single-pair
states can be of interest themselves.

We will be mostly concerned with the study of 
the anisotropic version of the model, the 
$t$-$J_z$ model,  for which the bulk stripe,
pairing, and Nagaoka states,
very similar to those of the isotropic $t$-$J$ model, have been
found in the past.\cite{sasha,pakwo} One can argue that  the $t$-$J_z$ model
should reproduce qualitatively the holes' ground state of the
isotropic $t$-$J$ model, although the details of such states might
differ. 
Moreover, for the bulk stripe phase  in this model
a very close {\it quantitative} agreement between the
analytical and numerical
methods have been demonstrated in our previous work.\cite{sasha}
Here we provide an extensive analysis of the stripes and other
excitations of the model (\ref{tj}) in the presence of
the boundaries.

\begin{figure}[t]
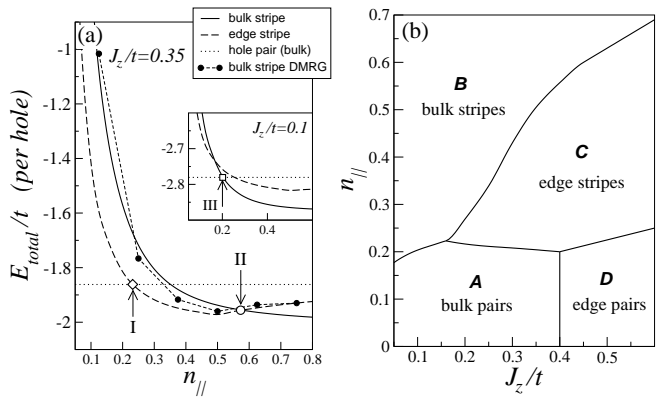
  
\includegraphics[angle=0,width=4.3cm]{Fig2a}\hskip 0.1cm
\includegraphics[angle=0,width=4.2cm]{Fig2b}
\caption{(a) Energy of the system per hole as a function of  hole
  density $n_{||}$ for the bulk and edge stripe, and the bulk pair state
  in the $t$-$J_z$ model (solid, dashed, and dotted,
  respectively). Black circles are the DMRG data for the 
bulk stripe,  Ref. \onlinecite{sasha}.  
$J_z/t=0.35$, and  $J_z/t=0.1$ (inset). Open symbols denote transitions: 
{\bf I} - pair to edge stripe, {\bf II} - edge to bulk stripe,
  {\bf III} - pair to bulk stripe.
(b) Constrained (see text)  phase diagram of the
$t$-$J_z$ system 
with an edge. ${\bm A}$ - bulk pairs, ${\bm B}$ - bulk stripes,
  ${\bm C}$ - edge stripes, ${\bm D}$ - edge pairs.} 
\label{energy}
\end{figure}

The stripe occurs due to the lowering of the
kinetic energy of holes at the 1D ADW.
 Although the
stripe is always beneficial from the point of view of kinetic energy,
there is a magnetic energy of the ADW, $\sim LJ_z$,
which one needs to compensate. Ignoring hole-hole
interaction the energy of the bulk stripe is given by:
\begin{eqnarray}
\label{total_E}
\frac{E_{total}}{N_h} = \frac{J_z}{2}\left(n_\parallel^{-1}-1\right)+
(\pi n_\|)^{-1}\int_0^{k_F}  \ E_{k}\ dk \ ,
\end{eqnarray}
where $E_{total}$ is relative to $E_0$ of the hole-free Ising AF
background, $k$ is a 1D wave-vector along the stripe. 
The first term in (\ref{total_E}) is the magnetic energy 
and the second term is the energy of the effective 1D band of holes 
filled up to the Fermi momentum $k_F=\pi n_\parallel$. The
quasiparticle energy $E_k$ is defined through the Dyson's equation:
%\begin{eqnarray}
%G_k^{-1}(E_{k}) = 
$E_{k} + 2 t \cos(k) - \Sigma(E_{k}) = 0$
% \ ,
%\label{gf}
%\end{eqnarray}   
where $\Sigma(\varepsilon)$ is the self-energy due to the string-like
hole motion away from stripe.  $\Sigma(\varepsilon)$
can be written in a compact form and Dyson equation 
can be solved numerically (see Ref. \onlinecite{sasha}). 
The bulk stripe energy 
%$E_{total}/N_h$ 
is shown in
Fig. \ref{energy}(a) as a function of $n_\parallel$ for two
representative $J_z/t$ values (solid lines). There is an 
excellent agreement of the theory with the
DMRG data for $J_z/t=0.35$ in an $11\times 8$ system (circles).
%----------------------------------------------------------------
%----------------------------------------------------------------
%----------------------------------------------------------------
\begin{figure*}
\includegraphics[angle=0,width=3cm]{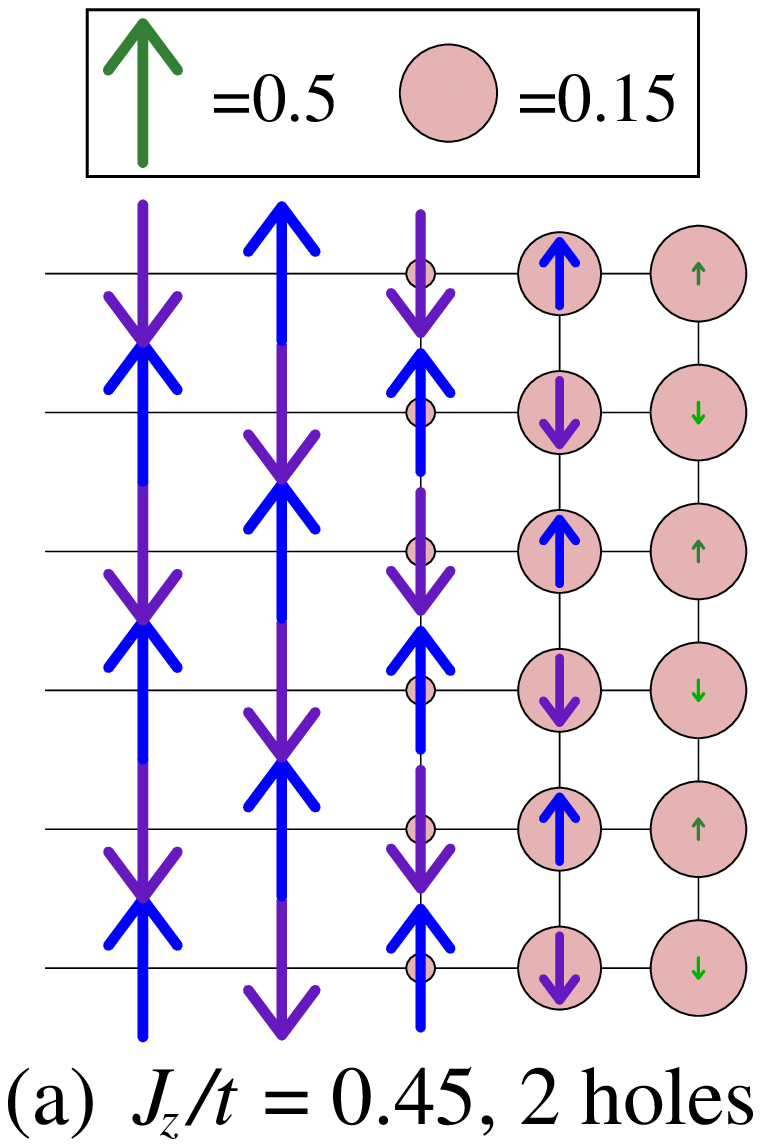}\hskip 0.5cm
\includegraphics[angle=0,width=3.5cm]{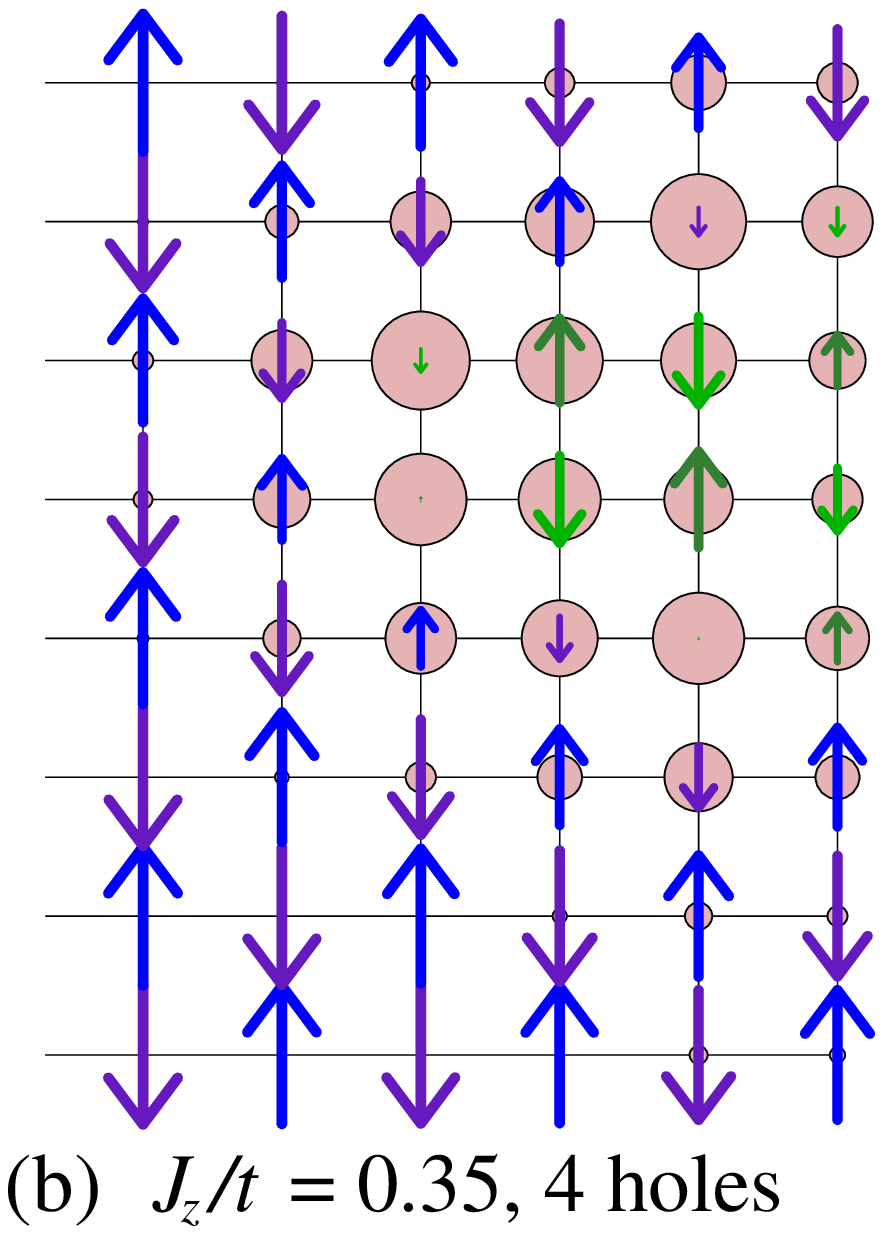}\hskip 0.5cm
\includegraphics[angle=0,width=10.2cm]{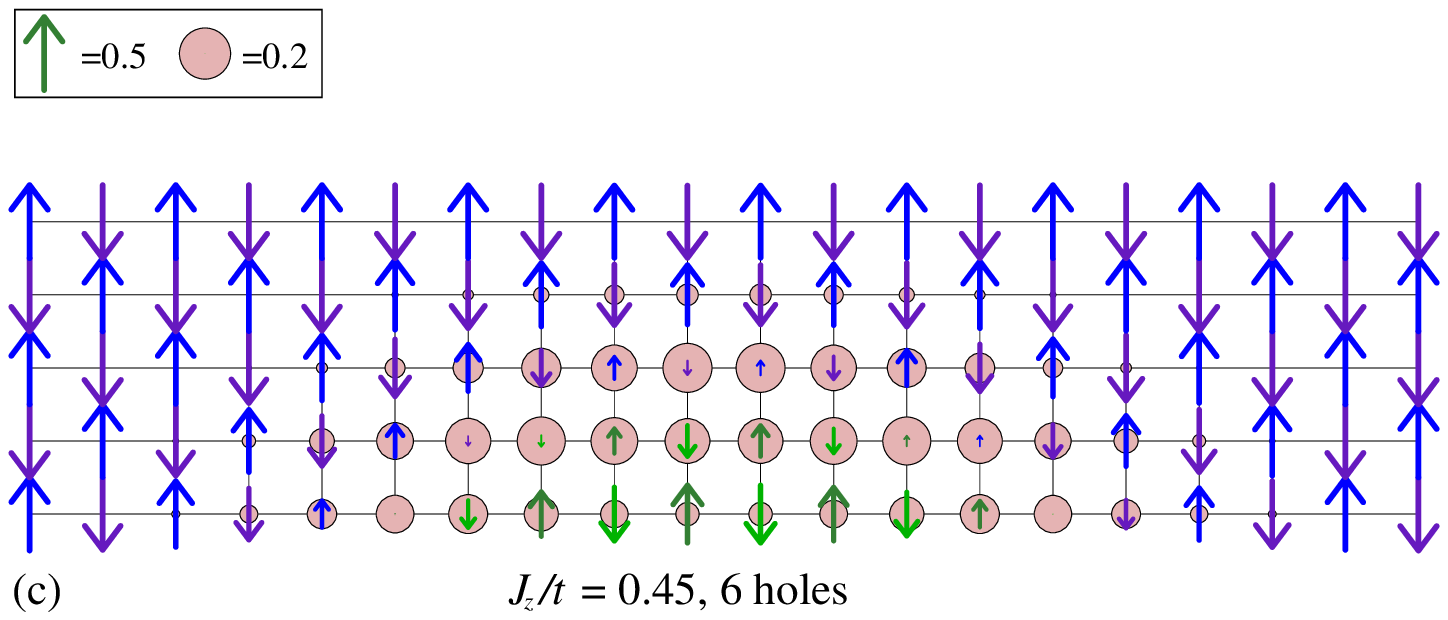}%\hskip 0.5cm
\caption{DMRG results for the edge stripes in the 
$t$-$J_z$ model in (a) $11\times 6$, (b)
  $11\times 8$, and (c) $20\times 5$ clusters. $J_z/t$ and the number of
  holes is as indicated. In (a) and (b) only the right half of the cluster
  containing all the holes is shown with the rightmost column being 
the open boundary. In (c) the bottom row is the ``free'' open boundary
(see text). The arrows represent the on-site
  value of $\langle S_i^z\rangle$ and the circles are for the 
hole density $\langle n_i\rangle$. Scales
for $\langle S_i^z\rangle$ and $\langle n_i\rangle$ in (b) and (c)
  are the same.}
\label{edge_forms}
\end{figure*} 
%----------------------------------------------------------------

This bulk stripe energy should be compared to the one of bulk
holes/pairs. In a homogeneous AF holes attract each other in a
bound state with the binding energy
$E_b$ ($\simeq -J_z/2$ for $J_z/t\gg 1$ and $\sim -J_z^3/t^2$ for
$J_z/t\ll 1$).\cite{pakwo} For the dilute system we assume that the
hole pairs do not overlap and $E^{pair}_{total}=N_h(E_{polaron}+E_b/2)$, 
shown by the straight lines in Fig. \ref{energy}(a).

In the presence of the boundary two types of new states are possible,
the edge pair and the edge stripe. Our
calculations within the  retraceable path approximation show that the
transition from the bulk pair to the edge pair occurs at $J_z/t\approx
0.4$. For the edge stripe the longitudinal
kinetic energy due to motion along the edge 
should remain the same as for the bulk  stripe
while the magnetic and transverse
kinetic energy change. At the stripe 
the hole excitations 
correspond to geometric kinks in the magnetic domain
wall.\cite{sasha} Therefore, placing the stripe
at the edge eliminates one-half of the broken bonds in
the ADW and the magnetic energy of the stripe is reduced to
$J_z/2\left(1/2 n_\|-1\right)$ per hole. 
The transverse kinetic energy is modified 
too, which leads to a different form of the self-energy
$\Sigma(\varepsilon)$ in the Dyson equation. As a
result the total energy of the edge stripe is given by the same 
expressions as in Eq. (\ref{total_E})  with the 
reduced magnetic
energy and different $\Sigma(\varepsilon)$.
The resulting
$E^{edge}_{total}$ are shown in Fig. \ref{energy}(a) by the dashed lines. 
One can clearly see the $1/n_\parallel$ behavior in the bulk and edge 
stripe energies
at low $n_\parallel$ where the magnetic energy dominates. For 
 $J_z/t=0.35$ the first transition, which occurs at
a lower $n_\parallel\approx 0.22$, is from the bulk pairs to the edge
stripe state. At higher density $n_\parallel\approx 0.58$ a transition
to the bulk stripe occurs. For the smaller value of $J_z/t=0.1$ the edge
stripe never materializes in the ground state and
the transition is only between the bulk stripe and bulk
pairs state (Fig. \ref{energy}(a), inset). Varying the
values of $J_z/t$ in such an analysis we obtain the ``constrained'' 
phase diagram in Fig. \ref{energy}(b).

At small $J_z/t$ the boundary is irrelevant and the competition
is between the bulk striped and paired states only. At low $n_\|$
stripes cannot be formed as $E_{mag}\propto J_z/n_\|$ is too high. At
intermediate $J_z/t$ the edge stripes form readily, while it is only at
larger $J_z/t$ when the edge pairs are formed. Note that the
pairs-to-stripes boundary in Fig. \ref{energy}(b) is only weakly
$J_z/t$-dependent. At larger $J_z/t$ the boundary 
is between the edge stripes and edge pairs. 
In principle, there can be  more
subtle varieties of the quasi-1D states at the edge, including
density-wave and others, which might be classified according to the
Luttinger liquid $g$-ology.\cite{luttinger}
Complicating such a classification is the bulk AF background breaking
spin 
%rotational
 symmetry.

% Because of our approximations, such as omission of the hole-hole
%interaction in the stripe state and independent pair approximation for
%the paired state, boundaries between different phases in
%Fig. \ref{energy}(a) may be approximate.
%In addition, a
As $J_z/t$ increases the edge state gets
strongly attached to the boundary and the energy difference between
the competing phases in this region becomes rather small. 
In that case, the qualitative
distinctions between the edge pairs and edge stripes, such as the
anti-phase shift of the AF order parameter across the stripe,
diminishes and one is not be able to
tell what bulk state a given edge state is associated with.
Thus, one can expect a crossover from the edge stripes to the edge
pairs at larger $J_z/t$. 

We have used DMRG to verify numerically the existence of these
states.\cite{dmrg} 
We have studied systems $L_x\times L_y$ up to 
$11 \times 8$ sites, with
periodic boundary conditions (BCs) in the $L_y$ direction 
and open BCs in $L_x$ direction (representing
the edges). No external field was applied to the open
boundaries.\cite{dmrg,sasha}  
Altogether, we have a close agreement between our analytical and
numerical results. All the principal
phases, bulk and edge stripe, and bulk and edge pair states
are identified. 
Because of the finite sizes of the modeled systems only a
discrete set of hole densities $n_\parallel$ is available, while 
$J_z/t$ can be varied continuously. 
We have studied the clusters
$11\times 4$ with one hole, $11\times 6$ with two holes, and $11\times
8$ with two and four holes. 
The number of states kept ranged from 1000 to 4000, with as many as 20
sweeps, allowing each run to converge nicely.
An evolution from the bulk
to the edge stripe was observed in  our $11\times 4$ ($n_\|=0.25$) and
$11\times 6$ ($n_\|=0.33$) clusters. 
There we start from the bulk stripe at small $J_z/t$ 
in the center of the cluster,\cite{sasha} which broadens and
becomes degenerate with the edge
stripe upon increase of $J_z/t$, and then forms the edge state,
see Fig. \ref{edge_forms}(a).
One can find remnants of the antiphase
 $\pi$-shift due to the ADW for the
edge stripe in Fig. \ref{edge_forms}(a), a clear signature that this
state remains quantum-mechanically connected to the bulk stripe. 
If, on the other hand, we start from the bulk paired state, ($11\times
8$, two holes) we have a transition to an edge state at around
$J_z/t=0.4$ (in agreement with the analytical calculation), which 
has no stripe-like magnetic correlation and we classify it as
the edge pair. 
Another distinction is the energy. One
can compare the energies per hole of the systems of different size 
with the same number of holes in which the edge pair and the edge stripe 
are formed ($11\times 6$ and $11\times 8$ with two holes).
Generally, such per hole energies are lower in the edge stripe state, 
which indicates that: (i) the edge stripe is distinct from the edge
pair, and that (ii) the edge state will be favored in the
thermodynamic limit. 
Upon further increasing of $J_z/t$ the energies of the edge states 
become degenerate as shown by such a comparison, which supports our 
 suggestion of the crossover from the edge stripe 
to the edge pair state at larger $J_z/t$.

Also, for the concentration $n_\|=0.25$ we 
observe no transition from paired states to stripes %or vice versa
as a function of $J_z/t$,
in a close agreement with a flat boundary between pairs and
stripes in our analytical results in Fig. \ref{energy}(b). 

Note here that we also found similar states in the
case of the fully isotropic $t$-$J$ model, with the phase boundaries
shifted to somewhat higher $J/t$ and $n_\parallel$. 
In particular, $11\times 8$
system with four holes ($n_\parallel =1/2$) 
definitely forms an edge stripe at $J/t=0.6$. The
same system exhibits a bulk stripe at $J/t=0.4$.
%, although a
%possibility of weak coupling to the edge cannot be completely ruled
%out. 
For $J/t=0.6$, on an $11\times 8$ cluster, we find the
following energies per hole for various dopings, in units of $t$:
one hole in the bulk -0.90(1), at the edge -0.94(1); 
one pair, bulk -1.03(1), edge -1.03(1);
$n_\parallel =1/2$ stripe, bulk -1.08(1), edge -~1.08(1).
Despite the degeneracies for the pair and stripe in the bulk versus the
edge, the edge pair and stripe appear quite stable in the DMRG run, and
strongly bound to the edge. $J/t=0.6$ appears to be quite close to the
transition between bulk and edge stripes and pairs.

Thus far our DMRG results for the low hole numbers have agreed
with the ``constrained'' analytical results in which only homogeneous 
hole states were taken into consideration. 
While this analysis can be directly relevant to certain
nano-scale structures doped with few holes, the
question remains which phases will survive in the thermodynamic
limit. 
Our energy-per-hole analysis shows that energy is always lower in
the stripe phase (bulk or edge). This implies an instability towards a
phase separation of a peculiar kind. That is, 
the system will tend to segregate into a hole empty region and
 a region where a stripe-like phase will be formed. Gas of pairs (edge or
bulk) will not be selected for the ground state. 
In fact, we observe an indication of such a phase separation already
in our $t$-$J_z$ 
$11\times 8$ cluster with four holes ($n_\|=0.5$). We start
at a lower $J_z/t$ with the ``regular'' bulk stripe which evolves into
an excellent edge stripe at larger $J_z/t$, similar to one in
Fig. \ref{edge_forms}(a). However, on its way to the edge it forms a
loop-like stripe object, with clear ADW, attached to the edge with
its ends, see Fig. \ref{edge_forms}(b).

We augmented our study of higher hole concentrations with the 
$20\times 5$ cluster with open BCs for all the edges and a strong
repulsive potential for holes (in the form of a chemical potential
$\mu=+t$) 
along the left, upper, and right edges
to create a preferential boundary and look for potential
edge states. Hole numbers up to six 
were studied. They organized themselves into an edge arc-like stripe, see
Fig. \ref{edge_forms}(c). Similar patterns have been produced by four
holes as well. Such states show clear signatures of
the ADW and are very stable. An ``internal'' hole concentration per
arc unit length can be estimated and is close to
$n_\|=1/2$. Thus, one can argue that the ground state of the 
semi-infinite $t-J$ system is likely to be composed of either
``straight'' or ``loop'' stripes (depending on a given $n_\|$), which
attach themselves to the boundary at some $J/t$.
Note, that nowhere in our study a mixed bulk-edge state was detected
and, as we discussed, the
$t$-$J_z$ or $t$-$J$ system with edges should belong to the class
described in our Fig. \ref{phd}(b).

In summary, we have studied the edge phase diagram of a 
strongly-correlated nano-structure
and found that it  consists of the bulk pairs, bulk stripes, 
and edge stripes. 
Using DMRG and analytical methods we have
shown that these edge stripe states are directly related to the bulk 
stripe states and have their origin in the non-trivial 
interplay between magnetic and kinetic energies in strongly-correlated
systems.  

We acknowledge illuminating conversations with J.~A.~Bonetti, P.~Mohanty, 
and D.~J. Van Harlingen.  
This research was supported by an award from Research
Corporation (ALC) and by NSF under grants DMR-0343790 (AHCN) and
DMR-0311843 (SRW).

\end{document}